\documentclass[twocolumn,showpacs,preprintnumbers,amsmath,amssymb,prb]{revtex4-1}
\usepackage{amsmath,amsfonts,bm,graphicx,color}
\usepackage[colorlinks=true,citecolor=black,linkcolor=black,urlcolor=blue,anchorcolor=black]{hyperref}
\newcommand{\degree}{\ensuremath{^\circ}}

\begin{document}

\title{Graphene on graphene antidot lattices: Electronic and transport properties}

\author{
S{\o}ren Schou Gregersen,
Jesper Goor Pedersen,
Stephen R. Power, and
Antti-Pekka Jauho
}
\affiliation{
Center for Nanostructured Graphene (CNG), Department of Micro- and Nanotechnology Engineering, Technical University of
Denmark, DK-2800 Kongens Lyngby, Denmark
}
\date{\today}

\begin{abstract}
	Graphene bilayer systems are known to exhibit a band gap when the layer symmetry is broken by applying a perpendicular electric field. The resulting band structure resembles that of a conventional semiconductor with a parabolic dispersion. Here, we introduce a novel bilayer graphene heterostructure, where single-layer graphene is placed on top of another layer of graphene with a regular lattice of antidots. We dub this class of graphene systems GOAL: graphene on graphene antidot lattice. By varying the structure geometry, band structure engineering can be performed to obtain linearly dispersing bands (with a high concomitant mobility), which nevertheless can be made gapped with the perpendicular field. We analyze the electronic structure and transport properties of various types of GOALs, and draw general conclusions about their properties to aid their design in experiments.
\end{abstract}


\maketitle

% ------------------------------
\section{Introduction}
% ------------------------------
The intrinsic properties of graphene, including ballistic transport, physical strength, and optical near-transparency, are very attractive for consumer electronics as well as for fundamental research platforms.~\cite{Geim2007,CastroNeto2009}
One of the main attractions of graphene is the prospect of manipulating its electronic properties and introducing a band gap, making the semimetal into a semiconductor as required for many electronic applications.~\cite{Xia2010,Schwierz2013,Schwierz2010}
As conventional potential barriers in graphene can exhibit Klein tunneling,~\cite{Geim2007,CastroNeto2009} much research has focused on finding methods to introduce a band gap into graphene.
Most proposals use structural modifications of graphene systems, such as nanoribbons, or superlattice structures imposed by periodic gating or strain.~\cite{Han2007,Nakada1996,Brey2006,Ezawa2006,Pereira2009a,Pereira2009,Pedersen2012a,Low2011,Neek-Amal2012}
More recent attempts use chemical modification through absorption or substitution.~\cite{Balog2010,Denis2010}
Periodic perforation of graphene sheets, to form so-called graphene antidot lattices (GAL), is of particular interest since theoretical predictions suggest the possibility of obtaining sizable band gaps.~\cite{Pedersen2008,Petersen2011,Pedersen2012,Gunst2011,Brun2014,Thomsen2014}
The band gaps of nanostructured graphene are however very sensitive to disorder and defects.~\cite{Yuan2013,Power2014}
Current nanostructure fabrication methods, e.g. block copolymer~\cite{Kim2010,Kim2012} or e-beam~\cite{Oberhuber2013,Giesbers2012,Eroms2009,Shen2008,Bai2010,Xu2013} lithography, will inevitably yield systems with a significant degree of disorder, especially near perforation edges.
Yet another emerging strategy towards altering the intrinsic behavior of graphene is to use structures composed of several 2D materials.
Bilayer graphene opens a band gap when an asymmetry is introduced between the two graphene layers.~\cite{Castro2010,Xia2010,LopesdosSantos2007,McCann2013,Ohta2006,Paez2014}
This is usually obtained by applying an electric field to create a potential difference between the top and bottom layers.
A transistor based on bilayer graphene has already been reported with a high on-off ratio $\sim 100$.~\cite{Xia2010}
Large areas of bilayer graphene can be fabricated, without etching, by mechanical exfoliation~\cite{Zhang2009} or by growth on a substrate~\cite{Ohta2006}, which reduces the risk of generating imperfections.
Unfortunately, most of these gapped or modified graphene systems lack the linear band structure of pristine graphene, e.g. bilayer graphene has a parabolic dispersion.~\cite{McCann2013,Ohta2006}
The implication of the parabolic bands is a lower mobility and thus degraded device performance.~\cite{Schwierz2010}
To overcome this, we propose the use of heterogeneous multi-layered structures. 
Bilayer superlattices have been studied in detail, with e.g. periodic potential barriers~\cite{Barbier2010}, and dual-layer antidot lattices~\cite{Kvashnin2014}.
A 1- or 2D potential modulation of the potential in bilayer graphene has even been predicted to yield linear dispersion.~\cite{Killi2011}
However, heterostructure bilayers composed of two different single-layer systems are not not widely studied.
Stacked heterostructures from multiple 2D materials created and held together only by van der Waals (vdW) forces~\cite{Geim2013} are particularly interesting as the interfaces may be kept clean from processing chemicals.

Previous studies have theoretically looked into single-layer doping in bilayer graphene,~\cite{Guillaume2012,Samuels2013,Mao2010,Collado2015} and experimentally single-sided oxygenation of bilayer graphene,~\cite{Felten2013} the latter of which reports electronic decoupling of one of the layers.
In this work we propose an all-carbon heterostructure that serves as a hybrid between single- and bilayer graphene.
It exhibits essentially linear bands at zero transverse bias while retaining the possibility of a bias-tunable band gap when dual-gating the top and bottom layers. 
The material is a bilayer heterostructure composed of a pristine graphene layer and a GAL layer, which we call \textbf{G}raphene \textbf{O}n (graphene) \textbf{A}ntidot \textbf{L}attice (GOAL).
We can hypothesize at least two methods in which a GOAL-based device could be realized experimentally, by either employing standard lithography~\cite{Oberhuber2013,Giesbers2012,Eroms2009,Shen2008,Bai2010,Xu2013} to etch the antidot pattern in only a single layer of bilayer graphene, or alternatively, by creating a sheet of GAL and then transferring pristine graphene on top using vdW stacking techniques.~\cite{Geim2013} 

The remainder of this paper is organized as follows. 
The atomic structure and the tight-binding model used for describing GOAL systems is introduced in Section \ref{sec:geom}. 
Section \ref{sec:prop} examines the properties of a representative sample of GOALs both with and without an applied bias. 
In Section \ref{sec:transport} the effects of different schemes for injecting current into and out of a GOAL device are addressed using two-lead transport simulations.
Finally, in Section \ref{sec:concl}, we discuss the implications of the investigated GOAL properties, the limitations of such systems and considerations relating to feasibility and application.

% ------------------------------
\section{Geometries and methods} \label{sec:geom}
% ------------------------------

\begin{figure}
\centering
\includegraphics{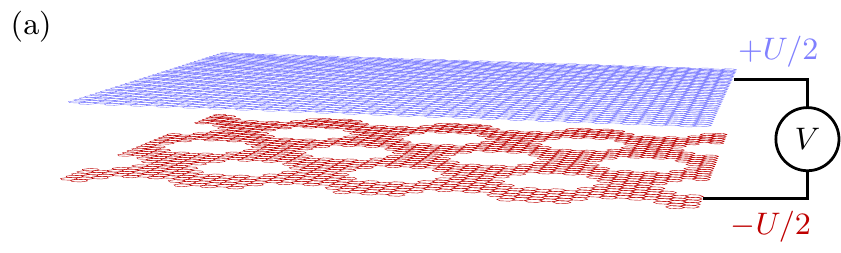}
\includegraphics{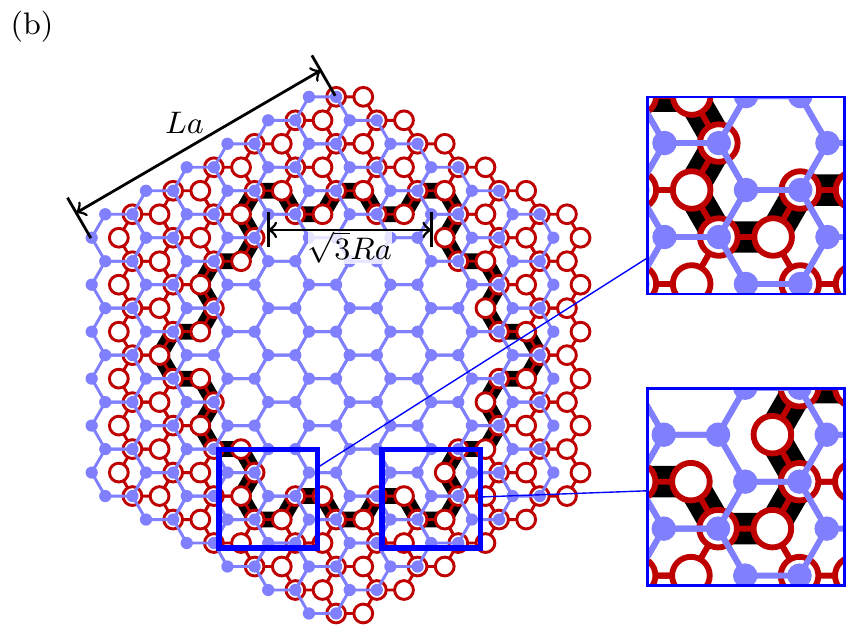}
\caption{
(a) Schematic illustration of the considered structures, consisting of a single graphene layer (blue) on top of a GAL layer (red), arranged in an AB stacking. 
(b) A closer view of the atomic structure of the Wigner--Seitz cell of a $\{L,R\}=\{6,2\}$ GOAL, with carbon atoms in the graphene (GAL) layer illustrated with blue filled circles (red open circles).
The integers $L$ and $R$ used for denoting a given geometry are illustrated and the antidot hole edge is highlighted by a black line. 
The GAL superlattice of the illustrated geometry is of the type that always has a band gap, as explained in the main text.
Zooms of two different corners of the antidot, corresponding to the thick blue outlines are shown on the right.
The corner-site in the bottom-left corner is a dimer, identified by the filled blue circle on top of an open red circle.
Conversely, the corner-site in the bottom-right corner is a non-dimer, identified by only either a filled blue or open red circle.
This gives rise to a $C_3$ symmetry, as discussed in the main text.
}
\label{fig:geom}
\end{figure}
We consider a heterostructure consisting of a single layer of pristine graphene on top of a layer of GAL, as illustrated schematically in Fig.~\ref{fig:geom}(a).
The twist angle between the layers greatly influences the electronic properties of bilayer graphene,~\cite{LopesdosSantos2007,Lee2011} and we expect the properties of the proposed GOAL structures to also depend on the angle between the two layers. 
However, for simplicity we focus in this paper on perfect Bernal (AB) stacking of the two layers. 
We discuss the possible influence of the angle in more detail in the final section of the paper.
Furthermore, experiments suggest the possibility of manually twisting the top layer until it `locks' into place at the Bernal stacking angle.~\cite{Dienwiebel2004}

Similar to the intricate edge dependence observed for graphene nanoribbons,~\cite{Nakada1996} the exact shape of the antidot greatly influences the electronic properties of isolated GALs.
In particular, extended regions of zigzag edges, which will generally be present for larger, circular holes, tend to induce quasi-localized states that significantly quench any present band gap.~\cite{Gunst2011, Brun2014} 
To simplify the analysis of the proposed structures we focus on hexagonal holes with armchair edges. 
Experimental techniques exist that tend to favor the creation of specific edge geometries.~\cite{Oberhuber2013,Jia2009,Pizzocchero2014, Xu2013}
In addition to the hole shape, the orientation of the GAL superlattice with respect to the pristine graphene lattice has a profound impact on the electronic properties.~\cite{Petersen2011, Brun2014}
The orientation of a superlattice may be defined by the vectors between two neighboring antidots $\mathbf{R}=n_1\mathbf{a}_1+n_2\mathbf{a}_2$, where $\mathbf{a}_1$ and $\mathbf{a}_2$ are the lattice vectors of pristine graphene. 
It has been shown that if $\mathrm{mod}(n_1-n_2,3)=0$ for any $\mathbf{R}$, the degeneracy at the Dirac point will break and a band gap is induced.~\cite{Odom2000,Petersen2011,Ouyang2011}
In this paper we consider GALs with two types of triangular superlattices: those with vectors parallel to carbon-carbon bonds which always induce a band gap, and those with vectors parallel to the pristine graphene lattice vectors which only induce gaps for a subset of superlattices. 
We only briefly discuss GOALs where the superlattice of the GAL layer is of the latter type, which we refer to as \emph{rotated} GOALs and \emph{rotated} GALs respectively, and focus mostly on the GAL superlattices for which band gaps are always present.
We demonstrate below that GOALs containing gapped GAL layers display similar properties regardless of the superlattice type, whereas GOALs with non-gapped GAL layers essentially behave as bilayer graphene with a renormalized Fermi velocity. 

The Wigner--Seitz cell of a specific GOAL is illustrated in Fig.~\ref{fig:geom}(b), where the red open circles represent the GAL layer atoms and the blue filled circles are the graphene layer atoms. 
To denote a given GOAL we use the notation $\{L,R\}$, where $La$ is the side length of the hexagonal unit cell, while $\sqrt{3}Ra$ is the side length of the hexagonal hole in the GAL layer, with $a=2.46$~{\AA} the graphene lattice constant.
We use $\{L,R\}_\mathrm{rot}$ to refer to GOALs in which the isolated GAL layer is of the rotated type, as discussed above.
Note that in this case, the Wigner--Seitz cell is not as shown in Fig.~\ref{fig:geom} but is rather in the shape of a rhombus with side length $La$.~\cite{Petersen2011} 
The condition for band gaps reads $L=3n+2$ where $n=0,1,...$ for isolated rotated GALs and within our model the other two-thirds of the rotated GALs are gap-less. 
The superlattice constant of a GOAL is $\Lambda=\sqrt{3}La$, while for a rotated GOAL it becomes $\Lambda_\mathrm{rot}=(L+1)a$.

In Bernal-stacked bilayer graphene there are four distinct sublattices, two in each layer. 
Within each layer we refer to these as dimer and non-dimer sites, and these sit directly above or below carbon sites (dimers) or the centers of hexagons (non-dimers) in the other layer.
These sites are illustrated in the right of Fig.~\ref{fig:geom}(b), where two of the antidot corners have been magnified.
It has been shown that the low energy properties of bilayer graphene are dominated by non-dimer sites, and can be described using an effective two-band model with parabolic bands touching at the Fermi energy.~\cite{McCann2013}
The introduction of the hole, forming the GAL layer of the GOAL system results in a higher number of sites from each sublattice in the graphene layer than in the GAL layer, but within our model maintains the sublattice symmetry within each individual layer.
The inter-layer asymmetry has important consequences when applying a bias across the layers, which we will discuss below in Sec. \ref{sec:prop:bias}.
Furthermore, the structures of GOALs no longer display a $60\degree$ rotational symmetry.
Neighboring corners of a hexagonal hole are now associated with sites from opposite sublattices, as can be seen on the right of Fig.~\ref{fig:geom}(b), reducing the $C_6$ symmetry of bilayer graphene to $C_3$.
Not all carbon sites in the graphene layer of a GOAL system are \emph{true} dimers or non-dimers, as the respective sites or hexagons below may have been removed by the holes.
However they still exhibit similar behavior to other sites in the same sublattice and we will thus collectively refer to them as dimers and non-dimers, respectively.

To calculate the electronic properties of the proposed structures, we use a nearest-neighbor tight-binding model.
The low-energy properties of single-layer graphene are quite accurately described by a model taking into account just the nearest-neighbor hopping term, $\gamma_0$. 
For bilayer graphene, additional inter-layer hopping terms need to be included. 
We consider the Slonczewski--Weiss--McClure model\cite{McCann2013} with the direct intra-layer hopping term $\gamma_1$ between AB dimers and the skew hopping terms $\gamma_3$ and $\gamma_4$ between dimers and non-dimers.
As we show below in Sec. \ref{sec:prop}, omitting the skew hopping terms has no qualitative impact on the results obtained. 
Therefore in most our calculations we disregard the skew hopping terms which are responsible for trigonal warping and electron-hole asymmetry in bilayer graphene.~\cite{McCann2013}
Furthermore, we do not include any on-site energy difference between dimer and non-dimer sites.~\cite{McCann2013}
The Hamiltonian then reads
\begin{align}
\mathbf{H} &= \sum_{i,j \in \{\mathrm{nn}\}} \gamma_0 \mathbf{c}_i  \mathbf{c}_j^\dagger + \sum_{i,j \in \{\mathrm{dimers}\}} \gamma_1 \mathbf{c}_i  \mathbf{c}_j^\dagger + h.c.
\end{align}
where $\{\mathrm{nn}\}$ is the collection of nearest neighbor pairs within each layer and $\{\mathrm{dimers}\}$ is the collection of dimer pairs.
We take $\gamma_0 = -3.16$~eV and $\gamma_1 = 0.381$~eV.\cite{McCann2013,Kuzmenko2009} 
An inter-layer bias $U$ (initially $U=0$) can be included via a shift $\pm U/2$ of the on-site energies on the GAL and the graphene layer, respectively. 
We define a positive bias to be one where the on-site energies of the graphene (GAL) layer are increased (decreased), as illustrated in Fig.~\ref{fig:geom}(a).

% ------------------------------
 \section{Electronic properties}\label{sec:prop}
% ------------------------------

\begin{figure*}
\centering
\includegraphics{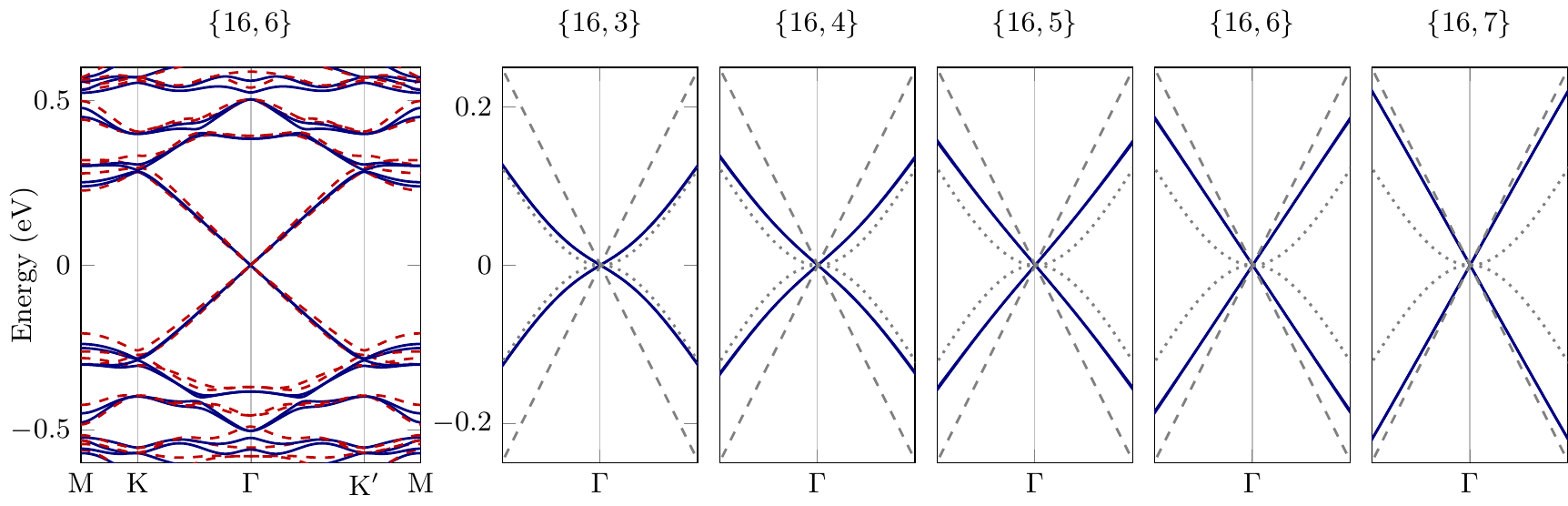}
\caption{Band structures of $\{16,R\}$ GOALs.
The left-most panel shows the full band structure within our model (solid blue lines), and for comparison the results obtained if skew scattering terms are included (red dashed lines). 
The right panels show a section of the band structure of GOALs near the $\Gamma$ point, for increasing antidot sizes, in solid lines.
Dashed gray lines show the corresponding single-layer graphene dispersion, while dotted gray lines illustrate the bilayer graphene dispersion.
}
\label{fig:bands}
\end {figure*}
We begin by examining the electronic band structures of some GOAL systems in the absence of a transverse bias. 
The left-most panel of Fig.~\ref{fig:bands} shows the band structure of a $\{16,6\}$ GOAL. 
The $\{16,R\}$ GOALs all contain GAL layers with a triangular superlattice, which in their isolated form are gapped for all $R$.
The solid lines show the band structure calculated with intra-layer and direct inter-layer hoppings only, whereas the dashed lines show the results obtained when including also the skew hopping terms, $\gamma_3 = -0.38$~eV and $\gamma_4=0.14$~eV.\cite{McCann2013,Kuzmenko2009}
The most striking features of the $\{16,6\}$ band structure are the linear bands near the Fermi energy, resembling the linear bands of single-layer graphene. 
The reduced Brillouin zone of the GOAL means that the $\mathrm{K}$ and $\mathrm{K}^\prime$ points of pristine graphene are folded onto the $\Gamma$ point. 
The most significant consequence of the skew hopping terms is to split the linear band into two linear bands with slightly different Fermi velocities.
The band splitting and the difference in Fermi velocities becomes more pronounced in cases near pristine bilayer graphene, where the antidot size is relatively small.
As we are mainly interested in a qualitative study of the proposed structures we disregard the skew hopping terms from hereon.

\begin{figure}
	\centering
	\includegraphics{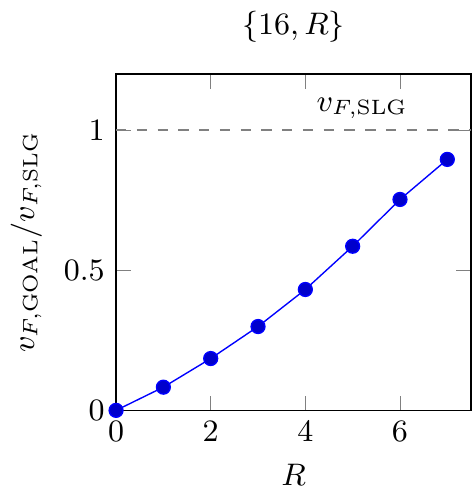}
	\caption{
		The Fermi velocity $v_{F,\mathrm{GOAL}}$ of $\{16,R\}$ GOALs as a function of $R$. 
		The $v_{F,\mathrm{GOAL}}$ is shown relative to the Fermi velocity of pristine graphene $v_{F,\mathrm{SLG}}$.
	}
	\label{fig:fermivel}
\end{figure}
To illustrate the transition from the parabolic bands of bilayer graphene to the linear bands of single-layer graphene as the antidot size is increased, we show in the right panels of Fig.~\ref{fig:bands} the dispersion relation near the $\Gamma$ point for the $\{16,R\}$ GOALs with increasing values of $R$. 
For comparison, the dashed (dotted) lines illustrate the pristine single-layer (bilayer) graphene dispersion, folded into the $\Gamma$ point. 
As the antidot size is increased, a transition from bilayer to single-layer-graphene-like (SLG-like) electronic properties is quite apparent, but with Fermi velocities which are slightly smaller than that of single-layer graphene.
This transition is also clear from Fig. \ref{fig:fermivel}, which plots the Fermi velocity of the $\{16,R\}$ GOALs at $E=0$ as a function of $R$.
The transition towards SLG-like bands does not occur via an ever increasing curvature of two parabolic bands touching at the Fermi energy. 
Instead, we always observe a region of linear bands for $R>0$, albeit the energy range in which the bands are linear is very narrow for small antidot sizes, and is accompanied by a strongly reduced Fermi velocity. 
Thus the low-energy band structure of GOAL can be considered as the crossing of two bands, similar to the case of single-layer graphene. 

\begin{figure}
\centering
\includegraphics{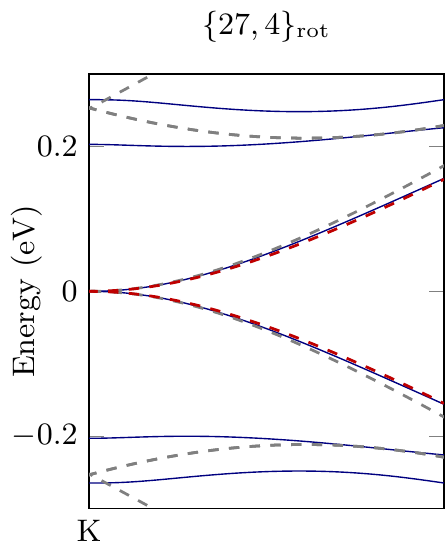}
\includegraphics{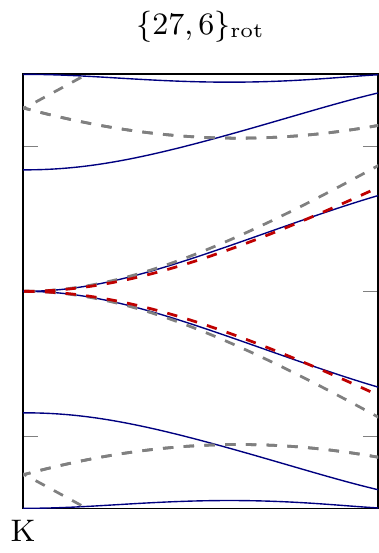}

\caption{
Band structures near the Dirac point of two $\{27,R\}_\mathrm{rot}$ GOALs with gapless GAL layers. 
The solid lines indicate the GOAL band structures, while the dashed gray lines are the band structure of pristine bilayer graphene. 
The dashed red lines show the bilayer graphene band structure with a renormalized Fermi velocity, as discussed in the main text.
}
\label{fig:bandsNonGapped}
\end {figure}
As the antidot size is increased more atoms are removed from the GAL layer and this leads to an effective reduction in the amount of bilayer graphene in the GOAL. 
We can quantify this via the relative area of bilayer graphene in the system, i.e. the ratio of the GAL and SLG layer areas, $f_\mathrm{BLG}=A_\mathrm{GAL}/A_\mathrm{SLG}=1-\tfrac{2\pi}{3\sqrt{3}}~\tfrac{R^2}{L^2}$. 
It is reasonable to ask whether the cause of the transition from parabolic to linear bands is simply caused by a reduction in $f_\mathrm{BLG}\rightarrow 0$ as $R$ is increased.
To determine whether this is indeed the case, we show in Fig.~\ref{fig:bandsNonGapped} the band structures near the Dirac point for two $\{27,R\}_\mathrm{rot}$ GOALs, which consist of gapless rotated GAL layers. 
The superlattice constants of the $\{27,R\}_\mathrm{rot}$ and the corresponding $\{16,R\}$ GOALs are roughly similar ($\Lambda/\Lambda_\mathrm{rot} \approx 1.01$) yielding very similar relative areas $f_\mathrm{BLG}$.
The band structures for the two $\{27,R\}_\mathrm{rot}$ GOALs are shown in solid lines together with those of bilayer graphene in dashed gray lines.
These rotated GOALs show a completely different dispersion, with no transition towards linear bands as the antidot size increases, even beyond the sizes shown in the figure. 
Despite having similar bilayer relative areas $f_\mathrm{BLG}$ to the GOALs considered in Fig.~\ref{fig:bands}, the band structures of the rotated GOALs remain parabolic and closely resemble that of pristine bilayer graphene.

We note that the isolated rotated GALs are gapless and that their band structures retain linear bands similar to pristine single-layer graphene, renormalized to a lower Fermi velocity.~\cite{Petersen2011}
This suggests that GOALs with gapless rotated GAL layers can be described by a model similar to that of bilayer graphene, but with a renormalized Fermi velocity.
The low-energy dispersion of bilayer graphene is well described in a continuum model,~\cite{McCann2013}
\begin{align}
E &= \pm 1/2 \gamma_1 \left[\sqrt{(1+4 \hbar^2 v_{F}^2 k^2 /\gamma_1^2} - 1\right]
\end{align}
where $v_{F}$ is the Fermi velocity of single-layer graphene.
To model the rotated GOAL we replace the Fermi velocity with the average Fermi velocity of the pristine graphene and renormalized GAL velocities, $\bar{v}_F$.
The results of this simple model are illustrated by red dashed lines in Fig.~\ref{fig:bandsNonGapped}, and indeed show quite good agreement with the full tight-binding results.
Interestingly, rotated GOALs with \emph{gapped} rotated GAL layers (e.g. $\{26,R\}_\mathrm{rot}$, not shown) display no qualitative difference from the regular GOALs with gapped non-rotated GAL layers.

\subsection{Distribution of states}\label{sec:prop:states}
\begin{figure}
\centering
\includegraphics{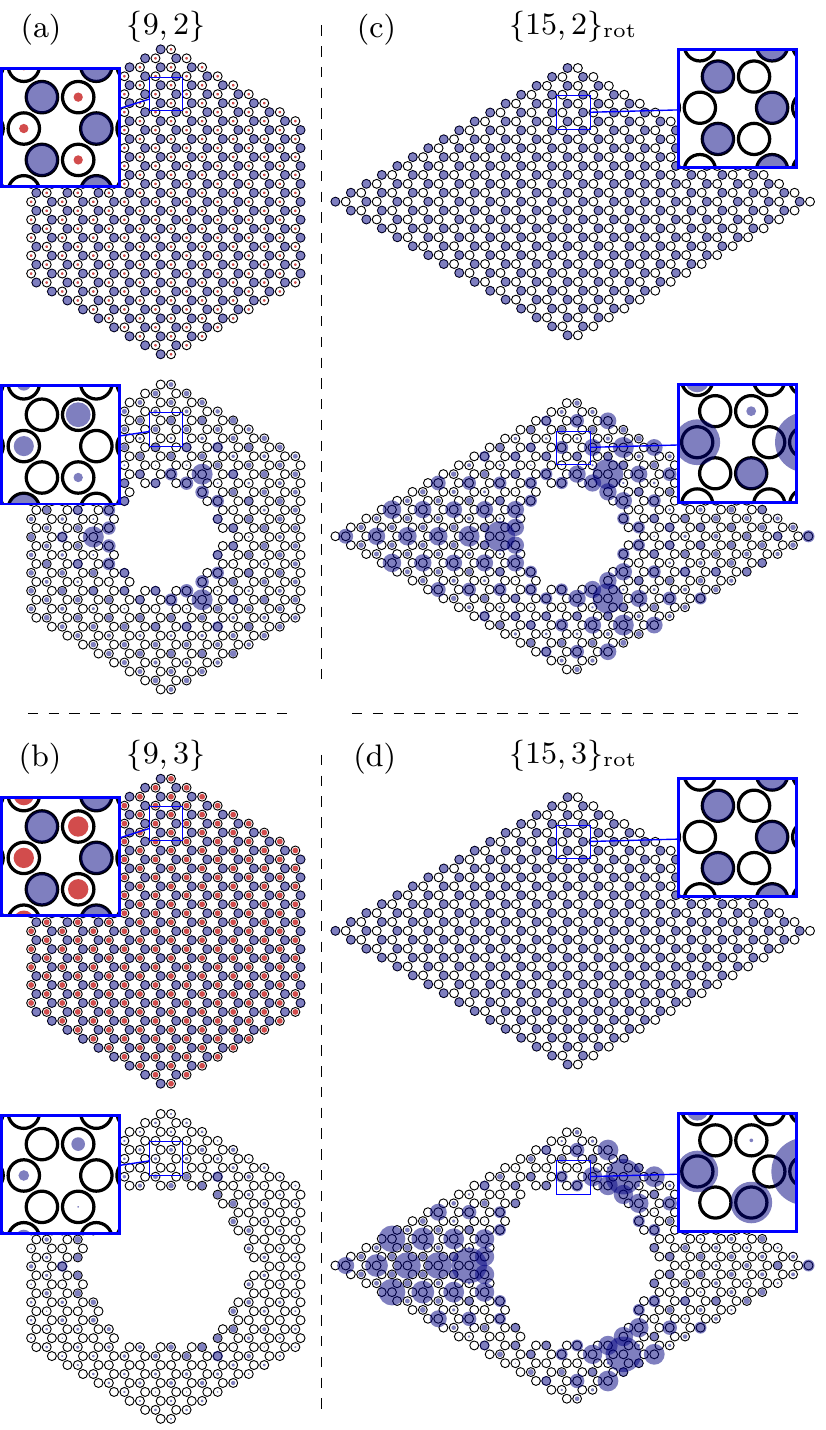}
\caption{
The projected density of states at the Fermi energy $E=0$. 
For the four systems considered, the PDOS of the two layers are displayed separately; the graphene layer above the GAL layer. 
The panels illustrate the PDOS of the $\{9,2\}$ GOAL (a), the $\{9,3\}$ GOAL (b), the $\{15,2\}_\mathrm{rot}$ GOAL (c), and the $\{15,3\}_\mathrm{rot}$ GOAL (d). 
The PDOS of dimer sites are illustrated by red filled circles and PDOS of non-dimer sites by blue filled circles. 
Their sizes represent the value of the PDOS relative to that of pristine single-layer graphene, shown by open circles.
Thus, if the PDOS is lower than that of pristine graphene the filled circles are smaller than the open circles and vice versa.
}
\label{fig:ldosExamples}
\end{figure}
The transition from parabolic to linear bands can thus not be explained entirely by the relative area of bilayer graphene, $f_\mathrm{BLG}$, in the GOAL system, but instead depends critically on the existence of a band gap in the isolated GAL layer.
To illustrate how the band gap of the GAL layer induces the SLG-like behavior in the combined system we show the projected density of states (PDOS) at the Fermi energy $E=0$ for each layer of the $\{9,2\}$ and $\{9,3\}$ GOALs in Fig.~\ref{fig:ldosExamples}(a) and (b). 
We will later discuss the differences in $\{15,R\}_\mathrm{rot}$ GOALs which consist of gapless GAL layers.
The properties illustrated by the $\{9,R\}$ GOALs are qualitatively similar to those of $\{16,R\}$.
The PDOS of the two layers are displayed separately, with the graphene layer above and the GAL layer below. 
Furthermore, the PDOS of dimers and non-dimers are illustrated by filled red and blue circles, respectively.
The size of the filled circles represents the value of the PDOS, which is normalized relative to that of pristine single-layer graphene shown by the open circles.
The PDOS of the $\{9,2\}$ and $\{9,3\}$ GOALs are illustrated in Fig.~\ref{fig:ldosExamples}(a) and (b), respectively. 
We recall that in the case of pristine single-layer (bilayer) graphene the Fermi energy density of states is equally distributed across all sites (all non-dimer sites). 
Examining first the graphene layers of the GOAL systems, we note that, unlike in bilayer graphene, there is a non-zero PDOS on dimer sites. 
Furthermore, this is equally distributed within the graphene layer, regardless of whether or not the sites are above another carbon site or above an antidot.
Comparing the $\{9,2\}$ and $\{9,3\}$ cases, we see that the PDOS on dimer sites in the graphene layer increases with the antidot size.
Meanwhile, the PDOS of the graphene layer non-dimers remains unchanged from that of single-layer graphene as the antidot size varies.
Interestingly, in the GAL layer dimer PDOS remains zero for all antidot sizes.
The PDOS of the non-dimer sites in the GAL layer displays a $C_3$ symmetry, yielding  a three-fold symmetric confinement around antidot corners associated with non-dimer sites.
Furthermore, the PDOS of the GAL layer non-dimers clearly decreases as the antidot size is increased.
The net result of these features is that, for large antidots, the PDOS eventually displays a distribution largely confined in the graphene layer.
This emerges from a decrease in the GAL layer non-dimer PDOS and an increase in that of the graphene layer dimer sites.

\begin{figure}
\centering
\includegraphics{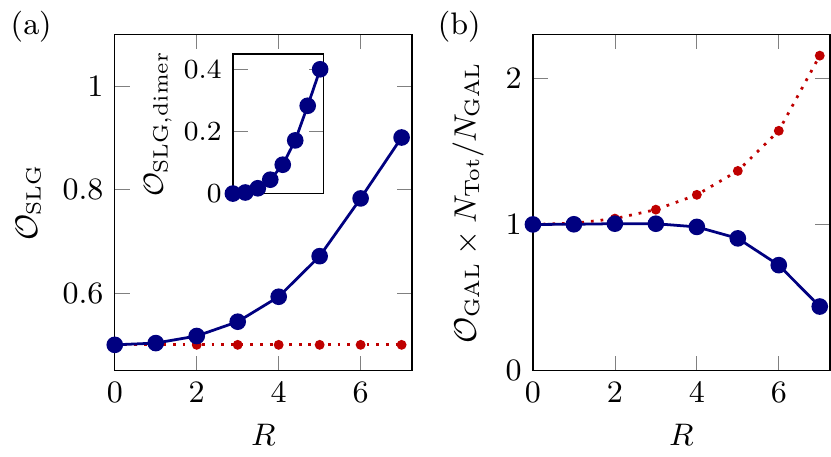}
\includegraphics{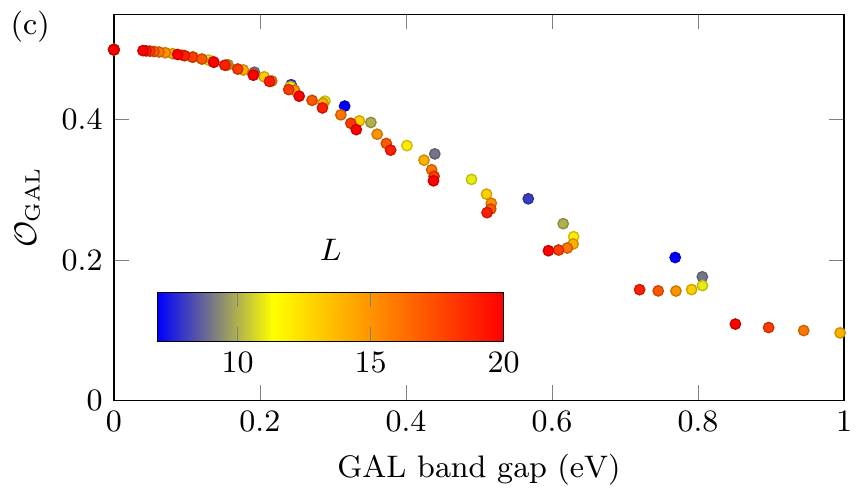}
\caption{
The integrated PDOS (overlap) of various GOALs.
(a) The overlap of the graphene layer for $\{16,R\}$ (solid lines) and $\{27,R\}_\mathrm{rot}$ (dashed lines) GOALs. 
The inset displays the dimer overlap in the graphene layer for the $\{16,R\}$ GOALs. 
The overlap of the non-dimers in the graphene layer, $\mathcal{O}_{\mathrm{SLG},\mathrm{non-dimers}}$, does not change.
(b) The relative overlap of the GAL layer for the $\{16,R\}$ (solid lines) and $\{27,R\}_\mathrm{rot}$ (dashed lines) GOALs. 
(c) The overlap with the GAL layer at the $\Gamma$ point versus the band gap of the isolated GAL layer for $\{L,R\}$ GOALs with $L\in [7;24]$ and valid $R$ within $ [0,L]$. 
The color of each dot indicates the value of $L$. 
}
\label{fig:ldos}
\end{figure}
We can illustrate these findings more clearly by considering the PDOS integrated over all sites within each of the layers, which we quantify via the overlap
\begin{align}
\mathcal{O}_i(E) \equiv \sum_{n}\sum_{m\in i} \left|c_{m}(E_n)\right|^2\delta\left(E-E_n\right),\label{eq:overlap}
\end{align}
where $c_m(E_n)$ is the expansion coefficient of the $n$'th eigenstate on to the $\pi$-orbital centered at the $m$'th atomic site, and where $i$ denotes the layer, $i \in \{\mathrm{GAL},\mathrm{SLG}\}$. 
A value of $\mathcal{O}_\mathrm{SLG}(E)=\mathcal{O}_\mathrm{GAL}(E)=\frac{1}{2}$ thus corresponds to an equal distribution of the eigenstates across both layers. 
The graphene layer localization at the Fermi energy is illustrated for $\{16,R\}$ GOALs in Fig.~\ref{fig:ldos}(a).
The solid line in the figure shows the graphene layer overlap as a function of antidot size.
As $R$ is increased the graphene layer overlap increases, i.e. the density of states become more confined in the graphene layer. 
The increased confinement is purely due to increased dimer PDOS, as apparent from the inset in Fig.~\ref{fig:ldos}(a) which displays the dimer overlap in the graphene layer, obtained by limiting the sum in Eq.~(\ref{eq:overlap}) to dimer sites, as a function of antidot size. 
The increased graphene layer localization could be due to a simple redistribution of the density of states on to the remaining sites, where the overlap is proportional to the number of sites in the particular layer.
We therefore consider the relative overlap $\mathcal{O}_i N_\mathrm{Tot}/N_i$, with $N_\mathrm{Tot}$ denoting the total number of carbon atoms with $R=0$, while $N_i$ the number of carbon atoms within the layer $i$. 
The value $\mathcal{O}_\mathrm{GAL} N_\mathrm{Tot}/N_\mathrm{GAL} = 1$ thus denotes a GOAL with layer overlaps proportional to the number of sites in that particular layer.
We show the relative overlap $\mathcal{O}_\mathrm{GAL} N_\mathrm{Tot}/N_\mathrm{GAL} $ of the $\{16,R\}$ GOALs in Fig.~\ref{fig:ldos}(b).
The solid line shows the relative overlap of the GAL layer as a function of the antidot size. 
The relative overlap is below unity for any non-zero $R$ and decreases with increasing antidot size.
Thus the GAL layer confinement decreases more quickly than a simple redistribution can account for, pushing the density of states even further into the graphene layer.
This transition from bilayer to single-layer confinement is critically dependent on the GAL band gap, and we therefore illustrate the GAL layer overlap for various $\{L,R\}$ GOALs as a function of the isolated GAL gap in Fig.~\ref{fig:ldos}(c).
Each GOAL is represented by a point colored by the value of $L$.
We find that the overlap in the GAL layer decreases with the GAL band gap in a largely one-to-one correlation, except at high GAL band gaps obtained through rather impractical antidot lattices, e.g. where the distance between antidots is only slightly larger than the antidot size.
As the GAL band gap increases states are pushed out of the GAL layer and into the graphene layer, effectively localizing the states in a single-layer yielding the SLG-like behavior.
This occurs, as we saw in Fig.~\ref{fig:ldosExamples}, via a transfer of states between the GAL layer non-dimer and graphene layer dimer sites as the antidot size, and thus the band gap, is increased.

To further illustrate the importance of the GAL band gap, we now consider the rotated GOALs which consist of gapless GAL layers and display a renormalized bilayer-like dispersion.
The PDOS at $E=0$ for the $\{15,2\}_\mathrm{rot}$ and $\{15,3\}_\mathrm{rot}$ GOALs are illustrated in Fig.~\ref{fig:ldosExamples}(c) and (d), respectively. 
The most notable feature in the rotated GOAL systems, as opposed to the non-rotated $\{9,R\}$ GOALs, is the zero PDOS of dimer sites in both layers of the rotated GOALs. 
The PDOS of the non-dimer sites in the graphene layer remains unaffected by the introduction of an antidot and the increasing of $R$. 
Therefore, the PDOS of the GAL layer non-dimer sites must increase.
This is more clearly seen in Fig.~\ref{fig:ldos}(a) where the graphene layer overlap of the $\{27,R\}_\mathrm{rot}$ GOALs is illustrated by the dotted red line. 
As the antidot size increases, no changes occur in the overlap of the graphene layer and hence also not in the overlap of the GAL layer.
In Fig.~\ref{fig:ldos}(b) we display the relative overlap of the GAL layer of the $\{27,R\}_\mathrm{rot}$  by the dotted red line.
In these rotated GOALs, the relative overlap increases above unity, corresponding to the redistribution of the PDOS onto the remaining non-dimer sites within the GAL layer.
This is also seen in the GAL layers of the $\{15,R\}_\mathrm{rot}$ GOALs shown in right panels of Fig.~\ref{fig:ldosExamples}, where the PDOS of the individual non-dimer sites has been significantly increased compared to the $\{9,R\}$ GOALs.
GOALs with gapless GAL layers do not push states into the graphene layer, but instead simply redistribute the density of states in the non-dimer sites of the GAL layer.
A low energy distribution of states amongst non-dimer sites only is a noted property of bilayer graphene, and confirms again the relation between the properties of rotated GOALs and those of the pristine bilayer. 
We limit the remainder of this paper to an investigation of the non-rotated GOALs, where the migration of states from the GAL to the graphene layer leads to an even distribution of states amongst the sublattices of the graphene layer, and thus to SLG-like behavior.

\subsection{Bias-tunable band gaps}\label{sec:prop:bias}
We now turn to biased structures. 
A potential difference between the layers induces a band gap in the case of pristine bilayer graphene, the size of which can be tuned by the bias voltage.~\cite{McCann2013, Nilsson2007, Castro2010, Ohta2006}
The potential $U$ can be created by a uniform electric field perpendicular to the two layers. 
In experimental systems the voltage difference $V$ is an induced quantity from the larger applied potential $V_\mathrm{ext}$ that due to screening and interlayer coupling is significantly reduced. 
For bilayer graphene the potential is uniform within the two layers and the induced voltage difference can be assumed linearly proportional to the applied voltage $V \propto V_\mathrm{ext}$, in which case currently $U$ has been predicted to realistically lie between $\pm$0.3~eV.~\cite{Nilsson2007} 
We note that in GOAL the edges will likely induce an inhomogeneous potential distribution. 
To find this distribution requires a self-consistent solution to the Poisson equation and band structure, a level of complication beyond the current scope.
We limit our model to include the bias via a uniformly distributed on-site energy shift $\pm U/2$ for the graphene and GAL layers respectively. 

\begin{figure}
\centering
\includegraphics{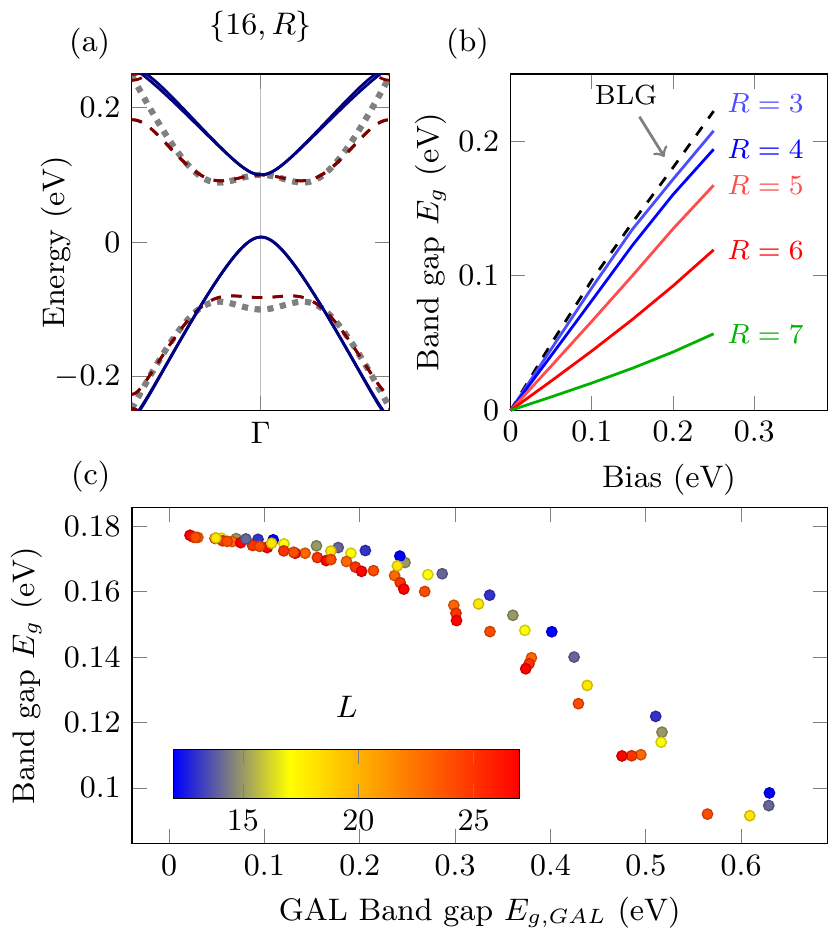}
\caption{
Band structures and gaps of biased various GOALs.
(a) Band structures for the $\{16,3\}$ (red, dashed) and $\{16,6\}$ GOALs (blue, solid) and pristine bilayer graphene (gray, dotted), with a bias $U=0.2$~eV applied across the layers. 
The bands resemble biased bilayer graphene, i.e. the ``Mexican hat" profile, for the small antidot $\{16,3\}$ and gapped single-layer graphene for the large antidot $\{16,6\}$ GOAL.
(b) Band gaps for $\{16,R\}$ GOALs with $R=3,4,5,6,7$ and an increasing bias. 
Note the near-linear dependence on the bias for all antidot sizes.
(c) The band gap of $\{L,R\}$ GOALs with a bias $U=0.2$~eV applied across the layers versus the isolated GAL layer gap, with $L\in [7;26]$ and valid $R$ within $ [0,L]$. 
The color of each dot indicates the value of $L$. 
}
\label{fig:biasBands}
\end{figure}
In a biased GOAL system, the inter-layer asymmetry of the on-site energies opens a band gap around the Dirac point.
We illustrate this in Fig.~\ref{fig:biasBands}(a) through the band structures of two biased $\{16,R\}$ GOALs at $U=0.2$~eV. 
In this figure, the bands of biased $\{16,3\}$ and $\{16,6\}$ GOALs are shown in dashed red and solid blue lines respectively, together with the bands of pristine biased bilayer graphene in dotted gray lines.
The band gap of biased $\{16,6\}$ GOAL is smaller than that of biased bilayer graphene or of the smaller antidot GOAL.
The change of the gap size is quantified in Fig.~\ref{fig:biasBands}(b) where we illustrate the band gaps of several biased $\{16,R\}$  GOALs as a function of $U$.
Each $\{16,R\}$ GOAL is shown as a solid line colored according to the value of $R$.
Additionally, the band gap of biased bilayer graphene is shown as a dashed line.
The band structures of the two biased $\{16,R\}$ GOALs in Fig.~\ref{fig:biasBands}(a) further display electron-hole asymmetry.
This arises due to the atomic imbalance between the two layers combined with the equal but opposite on-site energy shifts used to model the bias.
While the effect is minor in case of small antidots, for larger antidots the net energy shift caused by the imbalanced bias distribution yields a valence band shifted towards $E=0$.
We note also that the band structure of the biased $\{16,6\}$ GOAL resembles that of gapped graphene, identified by the absence of the ``Mexican hat" profile of biased bilayer graphene~\cite{McCann2013}. 
The absence of the flat profiles of biased bilayer graphene yields larger group velocities, which in turn is very attractive in fast electronic applications.
The transition between the bilayer graphene and gapped SLG-like dispersion is smooth, and similar to the zero-bias case can not be contributed solely to the reduced area $f_\mathrm{BLG}$. 
To illustrate this, we plot the biased GOAL band gap dependence on the isolated GAL gap for various $\{L,R\}$ GOALs in Fig.~\ref{fig:biasBands}(c) at $U=0.2$~eV, where each GOAL is represented by a point colored by the value of $L$.
The figure demonstrates clearly that an increase in the isolated GAL gap will cause a decrease of the biased GOAL band gap.
Although perhaps counterintuitive, this behavior is the direct result of GOALs with large band gap GAL layers exhibiting graphene layer confinement.
This effectively reduces the inter-layer asymmetry felt by the electronic states and reduces the band gap of the combined structure.
Fig.~\ref{fig:biasBands}(c) displays a clear correlation between the GAL band gap and the biased GOAL band gap, though it does display increased spreading as the GAL band gap is increased.
This spreading signifies an additional complication due to the uniform on-site energy shift $\pm U/2$ in the two asymmetric layers.
While the largest band gaps are found for GOAL systems whose unbiased electronic structure most closely resembles that of bilayer graphene, there is a range of $\{L,R\}$ values that yield both sizable band gaps and largely linear dispersion relations, e.g. the  $\{16,6\}$ shown here and also the $\{12,4\}$ case.
This presents the interesting possibility of combining high Fermi velocity electronic transport similar to single-layer graphene with a gate-controllable band gap. 

% % ------------------------------
 \section{Transport properties}\label{sec:transport}
% % ------------------------------

\begin{figure}
\centering
\includegraphics{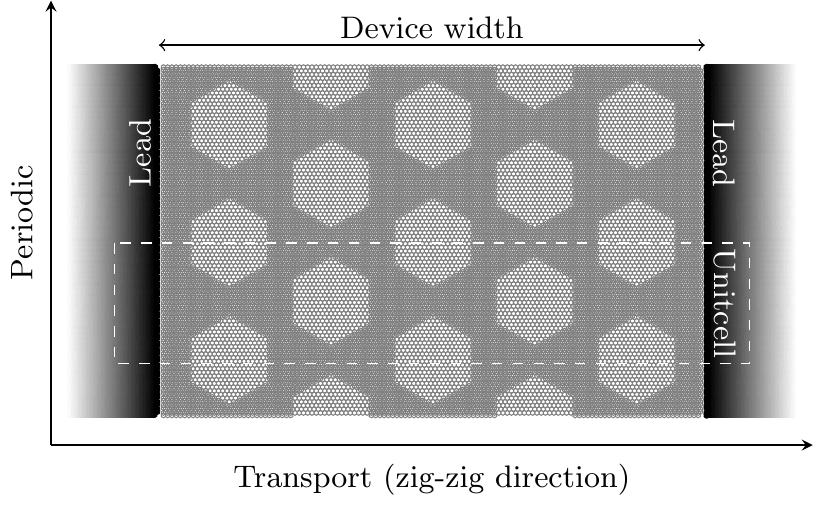}
\caption{
A schematic illustration of the GOAL device transport model. 
The incoming and outgoing leads (black), both of which are semi-infinite sheets of either single- or bilayer graphene, are coupled to a central GOAL device (gray). 
Bilayer leads are coupled to both layers of the GOAL device, while single-layer leads are coupled to either layer of the GOAL device.
The considered model is periodic in the transverse direction.
}
\label{fig:transport_setup}
\end{figure}
We mentioned two ways of experimentally fabricating GOAL devices; either by single-layer etching bilayer graphene or stacking a graphene sheet onto a GAL sheet. 
Most experimental transport measurements in bilayer graphene have been performed with top-contacts to inject current, and using dual-gates to control the inter-layer bias.~\cite{Oostinga2008,Du2008,Weitz2010}
With recent advances in side-contacts, first in single-layer graphene~\cite{Wang2013} and then in bilayer graphene~\cite{Maher2014}, there are now several ways of injecting current into a bilayer material such as GOAL.
The consequence of the choice of contacts has been studied for pristine bilayer graphene ribbons and flakes.~\cite{Gonzalez2010,Gonzalez2011}
To illustrate the consequences of the choice of contacts, we consider the electronic transport through a finite-width strip of GOAL.
To calculate the transport properties, we employ the Landauer-B\"uttiker formalism. 
The transport is calculated between two leads composed of either single- or bilayer graphene.
A schematic illustration of the transport model is shown in Fig.~\ref{fig:transport_setup}. 
In case of bilayer leads, these are connected to both the graphene and GAL layers, while single-layer leads are coupled to either the graphene or the GAL layer. 
Both the leads and the device are periodic in the transverse direction, and the unit cell used in calculations is outlined by the dashed rectangle.
We consider transport in the zig-zag-direction.
This yields a dense cross-section of antidots, effectively reducing the width of the GOAL device needed to represent large-width GOAL transport.~\cite{Gunst2011} 
Our calculations are performed on strips of GOAL with 7 antidots rows present along the transport direction.
This width yields a well defined transport gap in the isolated GAL layer.~\cite{Gunst2011}

With respect to the Landauer-B\"uttiker formula $G(E)=\frac{2e^2}{h} \mathcal{T}(E)$, the transmission $\mathcal{T}$ is determined using the Fisher-Lee relation which couples the transport to the Green's function of the full system.~\cite{Lewenkopf2013,Datta1995} 
The two leads are accounted for in the central device through the left ($\mathrm{L}$) and right ($\mathrm{R}$) self-energies $\boldsymbol{\Sigma}_\mathrm{L} $ and $\boldsymbol{\Sigma}_\mathrm{R}$.
The retarded Green's function at energy $E$ then reads
\begin{align}
\mathbf{G} (E) &= \left[ E + i\eta - \mathbf{H}_\mathrm{D}  - \boldsymbol{\Sigma}_\mathrm{L}(E) - \boldsymbol{\Sigma}_\mathrm{R}(E) \right]^{-1} \label{eq:greens}
\end{align}
where $\mathbf{H}_\mathrm{D}$ is the isolated Hamiltonian of the device region and $ i\eta$ is a small imaginary parameter needed for numerical stability.
Finally, the transmission is determined using the relation
\begin{align}
\mathcal{T}(E) = \mathrm{Tr}\left[ \boldsymbol{\Gamma}_\mathrm{R}(E) \mathbf{G}(E) \boldsymbol{\Gamma}_\mathrm{L}(E) \mathbf{G}^\dagger(E) \right]
\end{align}
where the $\boldsymbol{\Gamma}_{(\mathrm{L}/\mathrm{R})}(E) = - 2 \mathrm{Im}\left[\boldsymbol{\Sigma}_{(\mathrm{L}/\mathrm{R})}(E)\right]$ are the line widths for the respective leads. 
Bond currents through the device at specific energies are useful quantities in establishing how current flows through different parts of the device.~\cite{Lewenkopf2013}
The current between two neighboring sites $i$ and $j$ at the energy $E$ is~\cite{Cresti2003}
\begin{align}
I_{ij}(E) &=  \frac{4 e}{h}\mathrm{Im}\left[H_{ij}\left[\mathbf{G}(E) \boldsymbol{\Gamma}_L(E) \mathbf{G}^{*}(E)\right]_{ij}\right],
\end{align}
where $H_{ij} = [\mathbf{H}]_{ij}$ is the hopping term between the sites $i$ and $j$.
The transport calculations use both approximative recursive Green's function techniques to determine the lead self-energies and exact techniques for the device region to significantly speed up calculations, following Ref.~\citenum{Lewenkopf2013}.

\subsection{Transmission}
We consider two illustrative examples, the $\{16,3\}$ and $\{16,6\}$ GOALs.
From previous sections we recall that the $\{16,3\}$ and $\{16,6\}$ GOALs exhibit bilayer-like and single-layer-like dispersions, respectively. 
\begin{figure}
	\centering
	\includegraphics{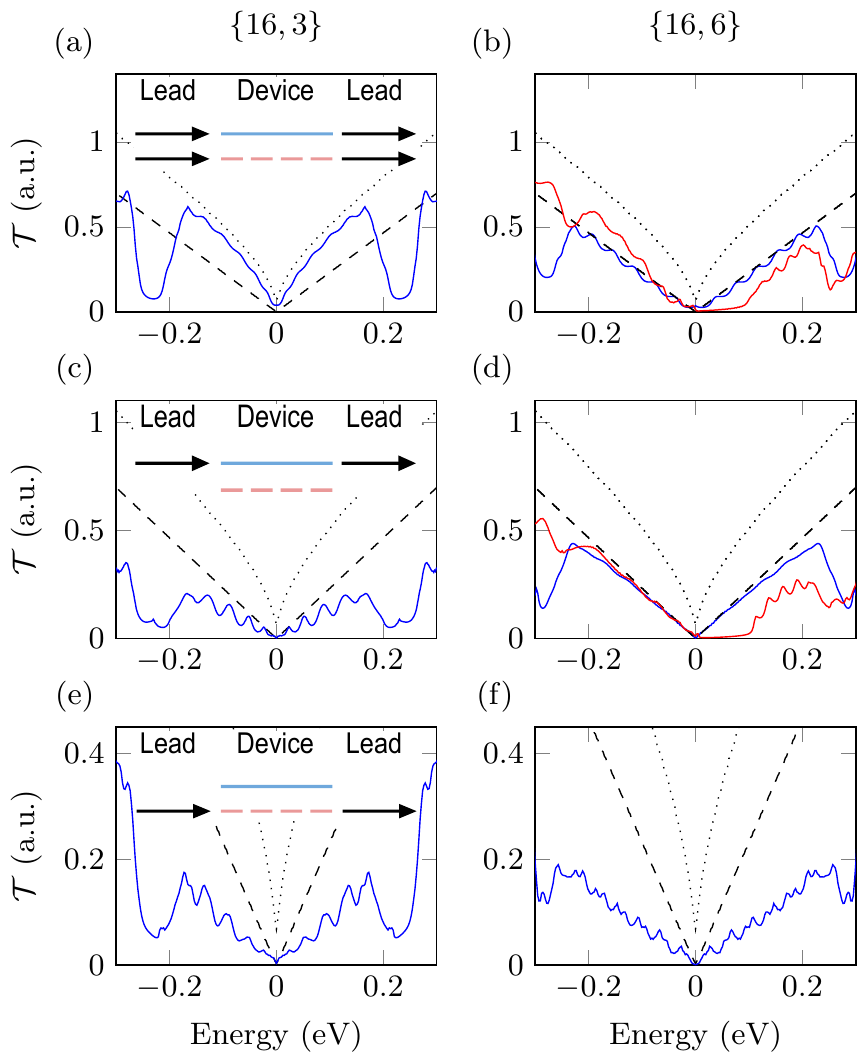}
	\caption{
		The transmission through $\{16,3\}$ and $\{16,6\}$ GOALs.
		The couplings are displayed in the insets of the left panels. 
		(a,b) Transport between two bilayer graphene leads through a central $\{16,3\}$ and $\{16,6\}$ GOAL device, respectively. 
		(c,d) transport between two single-layer graphene leads through a central $\{16,3\}$ and $\{16,6\}$ GOAL device coupling into the graphene layer, respectively. 
		(e,f) transport between two single-layer graphene leads through a central $\{16,3\}$ and $\{16,6\}$ GOAL device coupling into the GAL layer, respectively. 
		The central devices of (a,c,e) and (b,d,f) have the same widths receptively.
		The transmissions are displayed in solid blue lines along with pristine single- and bilayer graphene transmission, dashed black and dotted gray lines respectively.
		Additionally, (b) and (d) display transmission through a biased $\{16,6\}$ GOAL device coupled to bilayer graphene leads or single-layer graphene leads coupled to the graphene layer, respectively, in solid red lines.
	}
	\label{fig:transport}
\end{figure}
The transmissions between bilayer graphene leads connected to the $\{16,3\}$ and the $\{16,6\}$ GOAL devices are shown by solid blue lines in Fig.~\ref{fig:transport}(a) and (b), respectively. 
These transmissions are compared with pristine single- and bilayer graphene transmission, shown by dashed black and dotted gray lines, respectively.
Close to the Fermi energy, the transmission of the $\{16,3\}$ GOAL appears very similar to the pristine bilayer case, but with a slightly smaller magnitude.
This is consistent with the bilayer-like dispersion of the $\{16,3\}$ GOAL. 
In contrast, the $\{16,6\}$ GOAL transmission appears very similar to that of single-layer graphene. 
The qualitative transition from bilayer-like to single-layer-like transport behavior as a function of isolated GAL band gap is similar to that previously noted for the band dispersion.
Furthermore, an oscillatory behavior is observed which is particularly apparent for the $\{16,6\}$ transmission. 
By increasing the number of antidot rows beyond 7 (not shown) the transmissions yield an increased oscillation frequency, suggesting a Fabry-Perot like interference between scatterings at the lead-device interfaces.
The low transmission valleys just above $|E| \approx 0.2$~eV, which are present for both GOALs, appear at the end of the linear dispersion region and the onset of higher order bands.

The transmission between single-layer graphene leads coupled to the graphene layer of the GOALs is shown in Fig.~\ref{fig:transport}(c) and (d) (solid blue lines), compared again to pristine single- and bilayer graphene transmission (dashed black and dotted gray lines, respectively).
The transmission through the graphene layer of the $\{16,3\}$ GOAL is much lower than single-layer graphene transmission. 
This generally occurs for GOALs containing small-gap GAL layers due to wave mismatching, where the single-layer nature of the incoming wave is mismatched with the propagating bilayer waves in the GOAL device. 
We note that this also occurs in cases of bilayer graphene leads coupled to extremely large GAL gapped GOALs e.g. like $\{12,5\}$ where the incoming bilayer wave is mismatched with the single-layer nature of the GOAL device.
However, in the $\{16,6\}$ GOAL the layers are sufficiently decoupled to have single-layer-like propagating states, thus yielding a single-layer-like transmission.
Likewise, the Fabry-Perot oscillations have disappeared signifying lowered interface scattering, while they remain for the $\{16,3\}$ GOAL. 
The transmission between single-layer leads coupled to the GAL layer of $\{16,3\}$ and $\{16,6\}$ GOALs is shown in Fig.~\ref{fig:transport}(e) and (f), respectively. 
In this case the transmissions for both GOAL devices are lower than that of single-layer graphene.
The current must flow through either the GAL layer or couple in to and out of the graphene layer, which limits the transmission by the GAL band gap or the inter-layer couplings.

Finally, we consider the $\{16,6\}$ GOAL devices with an applied bias of $U=0.2$~eV.
The single layer and bilayer contact transmissions are illustrated in Fig.~\ref{fig:transport}(b) and (d) by red solid lines.
The band gap of the GOAL system forms a corresponding transport gap, effectively providing a SLG-like material with a tunable transport gap.
The optimal configuration for injecting current into a GOAL-based device should contact both layers, e.g. a side-contacted device.

\subsection{Bond currents}
In order to clarify the single-layer-like transport of GOALs, we now examine the bond currents in the systems studied above. 
We distinguish between in-plane and out-of-plane currents; currents flowing within either layer or currents flowing between the layers, respectively.
The model is the same as for the transmission illustrated in Fig.~\ref{fig:transport_setup}, where semi-infinite leads are coupled to a central GOAL device. 

\begin{figure}
\centering
\includegraphics{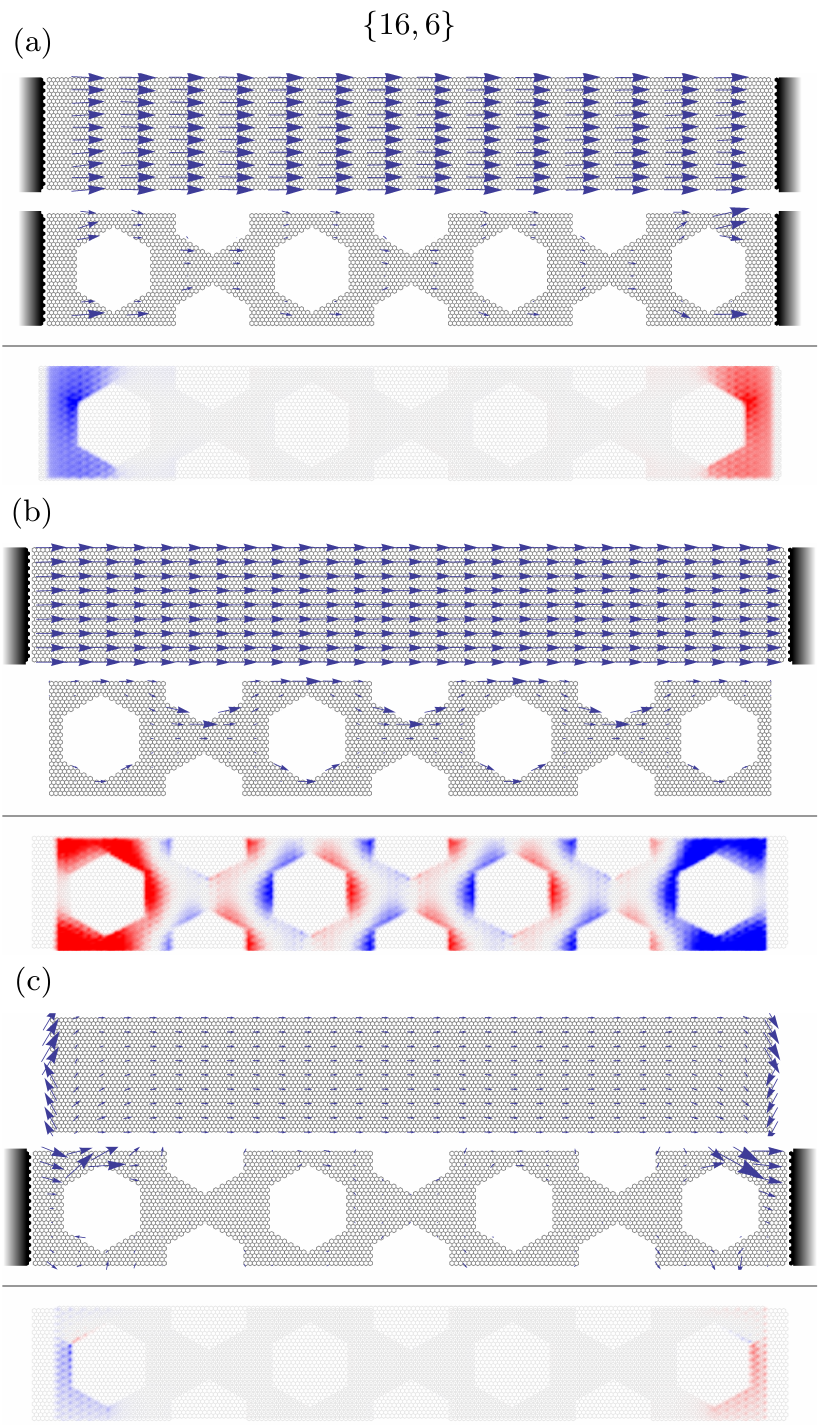}
\caption{
Current maps of GOAL transport devices.
In all panels, the in plane current maps are displayed separately, the graphene layer above the GAL layer, and the out of plane current maps are displayed below. The in plane currents are displayed as relative vectors scaled with the maximum in plane current within both layers. The out of plane currents are displayed as shaded areas colored according to the value, blue shading indicates current from the GAL layer into the graphene and red vice versa.
(a) The current maps of the $\{16,6\}$ GOAL device coupled to bilayer graphene leads.
(b) The current maps of the $\{16,6\}$ GOAL device coupled from the graphene layer to single-layer graphene leads.
(c) The current maps of the $\{16,6\}$ GOAL device coupled from the GAL layer to single-layer graphene leads.
}
\label{fig:currents}
\end{figure}
We consider the two cases where GOAL devices displayed transmissions similar to single-layer graphene, i.e. the $\{16,6\}$ GOAL device connected to either bilayer graphene leads or single-layer graphene leads which couple to the graphene layer only.
We illustrate current maps of the $\{16,6\}$ GOAL device at the energy $E=0.1$~eV in Fig.~\ref{fig:currents}.
In Fig.~\ref{fig:currents}(a) the currents of the $\{16,6\}$ GOAL device coupled to the bilayer leads is shown.
We plot the in-plane currents in each layer of the GOAL device separately, and show those of the graphene layer above those of the GAL layer.
These currents are displayed as vector maps, which are scaled relative to the maximum current in both layers.  
The most notable feature of the in-plane currents of the $\{16,6\}$ GOAL device with bilayer leads is the confinement of the current to the graphene layer throughout most of the device.
The out-of-plane current components are shown below the in-plane components as normalized color maps.
Blue shading represents for current flow from the GAL layer to the graphene layer, whilst red represents current from graphene layer to GAL layer.
This map displays a large current entering the graphene layer at the left interface and leaving at the right, yielding largely single-layer current transport.
The current within the GAL layer is not zero, and as the energy $E$ is increased the current within the GAL layer increases in magnitude.
The current thus becomes more and more bilayer-like as the energy of transport in increased, consistent with moving away from the band gap of the GAL layer.
In Fig.~\ref{fig:currents}(b) the bond currents in the $\{16,6\}$ GOAL device with a graphene layer connection to the single-layer leads are shown.
The in plane currents in this case also display noticeable confinement in the graphene layer.
However, in this case we observe that the in-plane current within the GAL layer is significantly larger.
The out-of-plane current map suggests the the current flows to the GAL layer near the left electrode and oscillates between the two layers near antidot edges, before returning to the graphene layer at the right electrode.
In both of these transport configurations, the current is largely confined to the graphene layer, yielding a transmission similar to, but slightly smaller than, single-layer transport.

Another interesting behavior occurs in the final case of single-layer leads connected to the GAL layer, illustrated in Fig.~\ref{fig:currents}(c).
In this case, the transport currents in a $\{16,6\}$ GOAL exhibit large edge currents within the graphene layer along the transverse (periodic) direction.
This behavior is a consequence of the high localization at every other corner in the hexagonal antidots, see Fig.~\ref{fig:ldosExamples}, such that the zigzag transport-direction will always scatter the current asymmetrically along the transverse direction.
If the same calculation is done along the armchair transport-direction, the scattering at the corners is symmetric and one finds much smaller and symmetric transverse currents.
Even though the transmission here is far smaller than single-layer graphene transport, the high transverse currents induced in the graphene layer suggest that interesting inter-layer transport couplings may be possible.
% 
% % ------------------------------
 \section{Discussion and Conclusion} \label{sec:concl}
% % ------------------------------
% Conclusion and discussion
In this work we have studied the electronic and transport properties of an all-carbon bilayer heterostructure consisting of a layer of pristine graphene atop a layer of nanostructured graphene.
In order to determine the general properties of such a heterostructure, we considered antidots as the ideal testbed, where structurally similar configurations yield entirely different single-layer properties.
These antidots were arranged into a triangular, or rotated triangular, superlattice orientation, yielding respectively gapped and gap-less antidot layers.
The electronic properties of the unbiased composite GOAL structures were seen to depend critically on the existence of this band gap in the isolated GAL layer.
A gapped GAL layer, regardless of superlattice orientation, will push electronic states into the graphene layer.
This is evident from the graphene layer confinement of the density of states, shown in Fig.~\ref{fig:ldos}(c), which increases with the GAL band gap.
As a consequence, the sublattice distribution of states seen in bilayer graphene is broken.
Instead we find an approximately even distribution of states between sublattices in the graphene layer, i.e. dimers as well as non-dimers.
Upon increasing the graphene layer confinement, the GOAL dispersion becomes linear near the Dirac point, and furthermore, the Fermi velocity increases until (at high GAL band gaps) it resembles that of pristine single-layer graphene.
Conversely, if the isolated GAL layer does not contain a gap, the GOAL composite retains a bilayer-like dispersion, except for a slight renormalization of the Fermi velocity.
The electronic state distribution in such GOALs is unchanged in the graphene layer, i.e. entirely located on non-dimers, while it is redistributed amongst the remaining sites in the GAL layer in a manner that conserves the pristine bilayer sublattice asymmetry. 
The dependence on the gap, and not directly the superlattice orientation or dimension, suggests a generality beyond this particular heterostructure.

Introducing an inter-layer bias to the GOALs with single-layer like dispersion induces band gaps smaller than those predicted for pristine bilayer graphene. The GOAL band gap size decreases as the band gap of its associated isolated GAL layer is increased. While GOALs with large-gap GAL layers have significantly reduced band gaps in the combined GOAL systems, specific GOAL structures were seen to exhibit both SLG-like dispersion and a sizable, tunable band gap.
Certain structures, such as the $\{16,6\}$ and $\{12,4\}$ GOALs, were identified which retained a high Fermi velocity in the unbiased case and sizable band gap in the biased case. 
Additionally, these GOAL systems when biased display gapped graphene-like bands, as opposed to the ``Mexican hat" shape bands of bilayer graphene.
The consequence is higher electron velocities than those in regular gapped bilayer graphene, which is of great interest in high-speed electronics.
Introducing a band gap in bilayer systems has been successfully done in experiments,~\cite{Ohta2006,Castro2007,Oostinga2008} and our results suggest a possibility of manipulating and fine tuning similar electronic behavior by nanostructuring of one of the layers.

In this work, we have limited our study to Bernal-stacked GOAL systems and to the most important coupling parameters, the intra-layer hopping $\gamma_0$ and inter-layer hopping $\gamma_1$.
Nonetheless, we expect more elaborate models to show the same qualitative results.
The inclusion of additional inter-layer couplings, responsible for electron-hole asymmetry and trigonal warping,~\cite{McCann2013} causes only a minor splitting of the bands near the Dirac point into two separate linear bands with slightly different Fermi velocities.
While this effect is more pronounced in GOALs with gap-less or smaller gap GAL layers, our focus is mainly on the more interesting single-layer-like GOALs with larger gap GAL layers.
It would however be very interesting to verify or modify these parameters through the use of \emph{ab initio} calculations specifically for GOALs.
Additionally, we employ a simple uniform potential distribution to describe the bias, which neglects edge effects that are likely to arise in these structures.
Given the intricate edge distribution of the density of states, the correct potential distribution may induce changes in the band edges of biased GOALs.
We also do not employ disorder or twisting of the GOAL systems.
In the case of disorder, this tends to decrease the band gap on an isolated GAL system.
The dispersion of the corresponding GOALs may exhibit transitions towards bilayer-like dispersion.
However, antidots with a hexagonal armchair shapes display higher stability against disorder than circular or hexagons with extended zigzag-edges.~\cite{Power2014}
By using experimental methods that prefer armchair edged shapes, this transition can be limited.
In case of twisting, models have been developed to illustrate what effect a small-angle twist has on the electronic properties in pristine twisted bilayer graphene.\cite{LopesdosSantos2007, LopesdosSantos2012}
Depending on the angle, the dispersion relations of twisted bilayers range from the parabolic bands of Bernal-stacked bilayer graphene to linear bands with a low Fermi velocity.~\cite{LopesdosSantos2012}
In the case of GOAL-based systems, the effect might be similar i.e. decreasing the Fermi velocity.
Furthermore, when the twisted bilayer graphene dispersion becomes linear the application of a perpendicular electric field is no longer guaranteed to open a band gap.~\cite{LopesdosSantos2007}
As such, the inclusion of a twist angle would require a more extensive study.

We have also studied transport properties including different contact configurations.
The transmission through GOALs exhibiting single-layer-like dispersion has approximatively the same magnitude as transmission through pristine graphene.
Furthermore, the current flow was largely confined to the graphene layer of the GOAL.
This follows from the electronic transport in pristine biased bilayer graphene, which depends greatly on the sublattice balances of the system.
The current density is greatest in the layer where the charge density is distributed equally across nondimers and dimers.~\cite{Paez2014}
The transport properties of GOALs also depend greatly on the type of contact to the device, similar to the case of pristine bilayer graphene.~\cite{Gonzalez2010,Gonzalez2011}
As the GOALs are bilayer materials, their propagating waves are also usually bilayer, albeit largely confined in the graphene layer.
This holds true except at very large GAL band gaps.
As such, GOALs display the highest transmission when coupling to bilayer graphene leads.
Unlike isolated GAL devices, the GAL layer of a GOAL device does not act as a barrier for transport.
Instead, the graphene-like transmission should be viewed as a result of \emph{mostly} single-layer confinement of the propagating states.
Coupling from single-layer leads, the mismatch between the incoming single-layer states and bilayer-like device states gives rise to increased interface scattering.
Except for very large GAL band gaps, this leads to transmissions below that of single-layer graphene.
The transmissions through GOAL devices with large-gapped GAL layers resemble that of SLG, suggesting single-layer-like propagation states.
In contrast to this, where single-layer leads connect only to the GAL layer the transmission is always low.
Both the lead/device wave mismatch and the current flow between the layers lead to the reduced transmission. 
Furthermore, in these cases the transport can display significant transverse currents within the graphene layer due to asymmetric scattering at hole edges.
For realistic devices, the best transmission is gained by injecting current into both layers, e.g. a side contact.

In this study we have demonstrated that the bilayer heterostructure can exhibit single-layer-like behavior similar to that of pristine graphene, while still allowing a tunable band gap.
The bilayers in this paper are seen to display a critical dependence on the band gap within the nanostructured layer. 
All results suggest that, as this band gap is increased the electronic states localize in the pristine layer, which yields monolayer behavior. 
From this, we expect that such a bilayer, with a gapless and a gapped layer, will transition from monolayer to bilayer behavior as the band gap within the gapped layer decreases.
Modifications which decrease such a gap may include structural defects, disorder and other imperfections, which in turn would lead to more bilayer-like behavior.
Many of the features discussed in this work may also be of relevance to other instances of 2D heterostructures where a metallic or semimetallic layer is coupled to a semiconducting or insulating layer.
We expect that in these cases a similar interplay between the electronic properties of the individual layers, and the redistribution of states when they are stacked, will determine the electronic and transport properties.
Such similar bilayer systems could include other forms of patterning of the nanostructured e.g. with dopants,~\cite{Jin2011,Guillaume2012,Samuels2013,Mao2010} absorbants,~\cite{Balog2010,Collado2015,Felten2013} or a Moir\'e potentials arising from coupling to a substrate.~\cite{Giovannetti2007}
Given the intense research currently underway in the field of nanostructured graphene, and the recent experimental progress in 2D heterostructure stacking, we believe that this type of composite system could bring interesting possibilities yet unseen in pristine graphene systems.

% 
% % ------------------------------
 \section{Acknowledgments} \label{sec:ackno}
% % ------------------------------
We thank Thomas Garm Pedersen for a fruitful discussion.
The Center for Nanostructured Graphene (CNG) is sponsored by the Danish Research Foundation, Project DNRF58.
The work by J.G.P. is financially supported by the Danish Council for Independent Research, FTP Grants No. 11-105204 and No. 11-120941.

\bibliography{GOAL}

\end{document}